\documentclass[aps,prd,twocolumn,showpacs,showkeys,preprintnumbers,amsmath,amssymb,amsfonts,nofootinbib]{revtex4}
\usepackage{bm}
\usepackage{enumerate}

\newcommand{\refb}[1]{(\ref{#1})}

\newcommand{\idxrangeb}[3]{\ensuremath{{#1}_{#2}\ldots{}{#1}_{#3}}}

\newcommand{\f}{L'}
\newcommand{\ord}{\textrm{order}}
\newcommand{\cchi}{{e_\perp}}
\newcommand{\ttt}{{e_\parallel}}
\newcommand{\h}{\perp}

\renewcommand\t[0]{
 \settowidth{\dimen7}{\mbox{\scriptsize{$\perp$}}}%
 \makebox[\dimen7][c]{\scriptsize{$\parallel$}}
}

\begin{document}

\author{M.~Cvitan}\email{mcvitan@phy.hr}
\author{S.~Pallua}\email{pallua@phy.hr}
\affiliation{
Department of Theoretical Physics,\\
Faculty of Natural Sciences and Mathematics,\\
University of Zagreb,\\
Bijeni\v{c}ka c.~32, pp.~331, 10000 Zagreb, Croatia}
\date{\today}
\preprint{Preprint ZTF 04-04}
\title{Conformal entropy for generalised gravity theories as a consequence of horizon properties}

\begin{abstract}
We show that microscopic entropy formula based on Virasoro algebra follows from properties of stationary Killing horizons for Lagrangians with arbitrary dependence on Riemann tensor. The properties used are consequence of regularity of invariants of Riemann tensor on the horizon.
Eventual generalisation of these results to Lagrangians with derivatives of Riemann tensor, as suggested by an example treated in the paper, relies on assuming regularity of invariants involving derivatives of Riemann tensor. This assumption however leads also to new interesting restrictions on metric functions near horizon.
\end{abstract}
%Gauss--Bonnet gravity;
%04.70.Dy Quantum aspects of black holes, evaporation, thermodynamics 
%11.25.Hf Conformal field theory, algebraic structures
%04.60.-m  Quantum gravity (see also 98.80.H Quantum cosmology)
%04.50.+h Gravity in more than four dimensions, Kaluza-Klein theory, unified field theories, alternative theories of gravity (see also 11.25.M Compactification and four-dimensional models), dilaton gravity

\maketitle

\section{Introduction}

One of the outstanding problems in gravity is to understand the nature of black hole entropy and in particular its microscopic interpretation. Except of being an important problem by itself, one can also hope that its solution would help to build the theory of quantum gravity. The problem of microscopic description of black hole entropy was approached by different methods like string theory which treated extremal black holes \cite{Strominger:1996sh}
or loop quantum gravity \cite{Ashtekar:1997yu,Dreyer:2004jy,Smolin:2004sx}. 

An interesting line of approach is based on conformal field theory and Virasoro algebra. 
One particular formulation for Einstein gravity was due to Solodukhin who reduced the problem of $D$-dimensional black holes to effective two dimensional theory with fixed boundary conditions on horizon. The effective theory was found to admit Virasoro algebra near horizon. Calculation of its 
central charge allows then to compute the entropy 
\cite{Solodukhin:1998tc}. The result was later generalised for $D$-dimensional Gauss--Bonnet gravity \cite{Cvitan:2002cs}. An independent formulation for two dimensional dilaton gravity and $D$-dimensional Einstein gravity is
due to Carlip 
\cite{Carlip:2002be,Carlip:1999cy}
who has shown that under certain simple assumptions on boundary 
conditions near black hole horizon one can identify Virasoro algebra as a subalgebra of 
algebra of diffeomorphisms. The
fixed boundary conditions give rise to central extensions of this 
algebra. The entropy is then calculated from Cardy formula 
\cite{Cardy:ie}
\begin{equation}\label{cardy}
S_{c}= 2\pi\sqrt{(\frac{c}{6}-4\Delta_{g})(\Delta-\frac{c}{24})}
\thickspace\textrm{.}
\end{equation}
Here $\Delta$ is the eigenvalue of Virasoro generator $L_0$ for the 
state we calculate the entropy and $\Delta_{g}$ is the smallest 
eigenvalue. 

In that way the well known Bekenstein--Hawking formula for Einstein gravity was reproduced. Explicitly
\begin{equation}\label{e58}
S= \frac{A}{4}.
\end{equation}
Here $S$ denotes black hole entropy and $A$ area of its horizon. Later these results have been generalised to include $D$-dimensional  Gauss--Bonnet gravity \cite{Cvitan:2002rh} and higher curvature Lagrangians \cite{Cvitan:2003vq}. For such case one can reproduce the generalised entropy formula \cite{Iyer:1994ys}
\begin{equation}\label{e60}    
S= \frac{\hat{A}}{4}= -2\pi\int_{\mathcal{H}}\hat{\epsilon }_{a_{1}\ldots a_{n-2}}E^{abcd}\eta_{ab}\eta_{cd}
.
\end{equation}
Here $\mathcal{H}$ is a cross section of the horizon, $\eta_{ab}$ denotes binormal to $\mathcal{H}$ and $\hat{\epsilon }_{a_{1}\ldots a_{n-2}}$ is the induced volume element on $\mathcal{H}$. The tensor $E^{abcd}$ is given by
\begin{equation}\label{e7}
E^{abcd} = \frac{\partial L}{\partial R_{abcd}}
.
\end{equation}

These derivations, however, included some additional plausible assumptions on boundary conditions near horizon. This includes in particular assumptions on behavior of the so called spatial derivatives assumed in Appendix A of Ref.~\cite{Carlip:1999cy}.

These assumptions have been even more crucial in the generalisations \cite{Cvitan:2002rh,Cvitan:2003vq} of original procedure. An important progress in understanding these assumptions can be done due to following observations \cite{Medved:2004tp} for stationary Killing horizons
\begin{enumerate}[I.]
\item The regularity of curvature invariants on horizon has strong implications on behaviour of metric functions near horizon
\item The transverse components of stress energy tensor have properties which suggest near-horizon conformal symmetry.
\end{enumerate} 
In fact using these results it was shown for $4$-dimensional Einstein gravity \cite{Cvitan:2004cj} that microscopic black hole entropy formula based on Virasoro algebra approach can be derived from properties of stationary Killing horizons. The above mentioned additional assumptions are shown to be fulfilled.

In this paper we would like to show that this is true not only for Einstein gravity but also for a generic class of Lagrangians which depend arbitrarily on Riemann tensor but do not depend on its covariant derivatives. Eventual exceptions which do not fall in this generic class will be defined more precisely in the text.

While in a previous case the results have been obtained by explicit calculations, this is not possible for generic case and thus we shall use a new method based on power counting.

We are interested in generalising the result from Einstein gravity to more general cases because  if that were possible it would indicate that near horizon conformal symmetries and corresponding Virasoro algebras are characteristic of any diffeomorphism invariant Lagrangian
and are independent of properties of particular Lagrangians and specific solutions. The additional interest in generalised Lagrangians is due to recent attempts to explain acceleration of universe by considering modifications of the Einstein--Hilbert action that become important only in regions of extremely low spacetime curvature \cite{Carroll:2003wy}. For more
 complete list of references see \cite{Carroll:2004de}.
In particular terms proposed to add to Einstein--Hilbert action have been of the type $R^{-n}$, $n>0$, and also inverse powers of $P=R_{ab}R^{ab}$ and $Q=R_{abcd}R^{abcd}$.
In this moment, a proof valid also for Lagrangians involving derivatives of Riemann tensor is still missing. However, we present an indication of possible similar results in Appendix \ref{appb} considering one specific example. One finds that additional restrictions on behaviour of metric functions near horizon are needed. However, these restrictions are very natural because it is found that they are consequence of regularity of invariants, this time involving derivatives of Riemann tensor. Thus, requesting regularity of invariants with derivatives of Riemann tensor gives new restrictions for metric functions near horizon in addition to those obtained by Medved, Martin and Visser \cite{Medved:2004tp}.

\section{Near horizon behaviour and derivation of the entropy}
As mentioned in the introduction we want explore if one can define Virasoro algebra at horizon its central charge and corresponding value for entropy for higher curvature Lagrangians of type
\begin{equation}\label{lgen}
L=L(g_{ab},R_{abcd}),
\end{equation} 
As explained in previous references \cite{Cvitan:2002rh}, the central charge is given with
\begin{equation}\label{e16}
\lbrace J[\xi_{1}], J[\xi_{2}] \rbrace^*=J[\lbrace\xi_1, \xi_2 \rbrace]+K[\xi_1, \xi_2]
\thickspace\textrm{,}
\end{equation}
where $\xi_1$, $\xi_2$ are diffeomorphisms generated by vector fields
\begin{equation}\label{e3}
\xi^a = T\chi^a + R\rho^a
\thickspace\textrm{,}
\end{equation}
where $\chi$ is Killing vector which is null on the horizon
\begin{equation}\label{e4}
\chi^2 = 0,
\end{equation}
and $\rho$ is defined with
\begin{equation}\label{e5}
\nabla_a \chi^2 = -2 \kappa \rho_a.
\end{equation} 

Diffeomorphism functions $T$ and $R$ are restricted with conditions
\begin{equation}
R = -\frac{1}{\kappa}\frac{\chi^2}{\rho^2}\nabla_\chi T,
\end{equation} 
\begin{equation}
\rho^a \nabla_a T = 0.
\end{equation}
In such a way surface $\chi^2 = 0$ will remain fixed under these diffeomorphisms and also
\begin{equation}
\frac{\delta \chi^2}{\chi^2} = 0,
\end{equation} 
will be valid. One more condition on diffeomorphisms is required
\begin{equation}\label{e22.3}
\delta \int_{\partial C}\hat{\epsilon}(\hat{\kappa}-\frac{\rho}{|\chi|}\kappa)=0
.
\end{equation}
Here, $\hat{\kappa} ^2=-\frac{a^2}{\chi^2}$ and $a^{a}=\chi^b\nabla _b\chi^a$
is the acceleration of an orbit of $\chi^a$. For more complete definition of diffeomorphisms see \cite{Carlip:1999cy}. This last condition leads to orthogonality relations for one parameter group of diffeomorphisms.

Now, Dirac bracket of boundary terms $J[\xi]$ in Hamiltonian
$ \lbrace J[\xi_{1}], J[\xi_{2}]\rbrace^*$
is given with (see Eq.~27 of \cite{Cvitan:2002rh})
\begin{eqnarray}\label{e6}
&&\lbrace J[\xi_{1}], J[\xi_{2}] \rbrace^* = \int_{\mathcal{H}} 
\epsilon_{apa_{1}\cdots a_{n-2}} \nonumber\\
&&\qquad\left\lbrace 2 \left(\, \xi_{2}^{p} 
E^{abcd}\nabla _{d} \delta_{1}g_{bc}-\xi_{1}^{p}\nabla_{d} 
E^{abcd}\delta_{2}g_{bc}-(1\leftrightarrow 2) \right) \right. 
\nonumber \\ 
&&\qquad \left. -\, \xi_{2}\cdot 
(\xi_{1}\cdot \mathbf{L})\right\rbrace  .
\end{eqnarray}

The information about Lagrangian is given with quantities $E^{abcd}$.
We introduce the following abbreviations
\begin{equation}
X^{(12)}_{abcd} = \xi_{1}^{p}\eta_{ap} 
\nabla_d\delta_2g_{bc}
- (1\leftrightarrow 2)
\thickspace\textrm{,}
\end{equation} 
\begin{equation}
\tilde{X}^{(12)}_{abc} = \xi_{1}^{p}\eta_{ap}
\delta_2g_{bc}- (1\leftrightarrow 2)
\thickspace\textrm{.}
\end{equation} 
In such a way \refb{e6} becomes

\begin{eqnarray}\nonumber
{\left\lbrace J[\xi_1], J[\xi_2] \right\rbrace}^*=
-\int_{\mathcal{H}} &\hat{\bm{\epsilon}}&
\left\lbrace 2\left(
  X^{(12)}_{abcd}E^{abcd}
  -\tilde{X}^{(12)}_{abc}\nabla_d E^{abcd}
\right)\right.\\
&-& \left.
\vphantom{\left(X^{(12)}_{abcd}\right)}
\xi_2^a\xi_1^b\eta_{ab}L\right\rbrace
\label{e31b}
\thickspace\textrm{.}
\end{eqnarray}

We are interested to evaluate this expression on horizon. The third term is immediately seen to be zero due to \refb{e3}, \refb{e4}, \refb{e5} and regularity of Lagrangian on horizon.
It will be shown that first two terms are given as follows
\begin{widetext}
\begin{equation}\label{lim1}
\lim_{n \rightarrow 0}\left(X^{(12)}_{abcd}E^{abcd}\right) = \lim_{n \rightarrow 0}\left(-\frac{1}{4}\eta_{ab}\eta_{cd}E^{abcd}
\left[(\frac{1}{\kappa}T_1\dddot{T}_2-2\kappa T_1\dot{T}_2)
- (1\leftrightarrow 2)
\right]
\right) 
\thickspace\textrm{,}
\end{equation}
\end{widetext}
\begin{equation}\label{lim2}
\lim_{n \rightarrow 0}\left(\tilde{X}^{(12)}_{abc}\nabla_d E^{abcd}\right) = 0
\thickspace\textrm{.}
\end{equation}
This was shown in \cite{Cvitan:2004cj} for Einstein Lagrangian and for Lagrangians including quadratic terms in curvature. Here we want to extend it to Lagrangians of general form given with \refb{lgen}.

The main properties which we shall use in this paper will be the properties of stationary horizon. In particular we shall use results of \cite{Medved:2004tp} where it was shown that absence of curvature singularities implies explicit restrictions on Taylor series of metric functions near horizon. For basis we use two Killing vectors of axially symmetric black holes
\begin{equation}\label{e1}
t^a = \left(\frac{\partial}{\partial t}\right)^a
\thickspace\textrm{,}\qquad
\phi^a = \left(\frac{\partial}{\partial \phi}\right)^a
\thickspace\textrm{,}
\end{equation} 
with corresponding coordinates $t$, $\phi$. In addition in the equal time hypersurface we choose Gauss normal coordinate $n$ ($n=0$ on the horizon) and the remaining coordinate $z$ such that the metric has the form
\begin{eqnarray}\label{e2}
d s^2 & = & - N(n,z)^2 d t^2 + g_{\phi\phi}(n,z) 
\left( d\phi - \omega(n,z) d t \right)^2 
\\\nonumber 
&+& d n^2+ g_{zz}(n,z) d z^2
\thickspace\textrm{.}
\end{eqnarray}

The mentioned properties imply that near horizon metric coefficients have following Taylor expansions

\begin{eqnarray}\label{mt}
N(n,z)&=& \kappa n + \frac{1}{3!} \kappa_2(z) n^3 +  O(n^4)\\\nonumber
g_{\phi\phi}(n,z) &=&  g_{H\phi\phi}(z) + 
\frac{1}{2} g_{2\phi\phi}(z) n^2 + O(n^3)\\\nonumber
g_{zz}(n,z) &=&  g_{Hzz}(z) + 
\frac{1}{2} g_{2zz}(z) n^2 + O(n^3)\\\nonumber
\omega(n,z) &=& \Omega_H + \frac{1}{2} \omega_2(z) n^2 +  O(n^3)
\thickspace\textrm{.}
\end{eqnarray} 

In the following we will use the basis ${e_\mu}^a$ where 
${e_1}^a \equiv \chi^a$,
${e_2}^a \equiv \rho^a$,
${e_3}^a \equiv \phi^a$,
${e_4}^a \equiv z^a$.

Leading terms of nonvanishing products of basis vectors are
\begin{eqnarray}\label{CT1}
\chi\cdot\chi &=& -\kappa^2 n^2 + O(n^4)\\\nonumber
\chi\cdot\phi &=& -\frac{1}{2} {g_{H\phi \phi }(z)}\,
\omega_2(z)\,n^2 
+ O(n^3)
\\\nonumber
\phi\cdot\phi &=& g_{H\phi\phi}(z) + O(n^2)\\\nonumber
\rho\cdot\rho &=& \kappa^2 n^2 + O(n^4)\\\nonumber
\rho\cdot z &=& O(n^4)\\\nonumber
z\cdot z &=& g_{Hzz}(z) + O(n^2)
\thickspace\textrm{,}
\end{eqnarray} 
and all other products are zero
\begin{equation}\label{CT2}
\chi\cdot\rho =
\chi\cdot z =
\phi\cdot\rho =
\phi\cdot z = 0
\thickspace\textrm{.}
\end{equation} 

It will be convenient to use $\cchi^a$ for $\chi^a$ or $\rho^a$ when equations hold for both $\chi^a$ and $\rho^a$, and similarly $\ttt^a$ for $z^a$ and $\phi^a$.

In the evaluation of \refb{lim1} and \refb{lim2} it is important to realise that tensors $X^{(12)}_{abcd}$ and $\tilde{X}^{(12)}_{abc}$ depend only on details of black hole and its symmetry properties but their form does not depend on the form of the Lagrangian. Also the derivation of \refb{lim1} depends only on symmetry properties of tensor $E$ and not on its particular form. For that reason Eq.~\refb{lim1} can be calculated as in \cite{Cvitan:2004cj}. 

However the proof of statement \refb{lim2} for Einstein gravity and Lagrangians quadratic in Riemann tensor was based on explicit calculations. These are of course not possible for generic class of Lagrangians of the type \refb{lgen}. Thus we need new approach.

The derivation in this case will be based on properties \refb{mt} of metric functions near horizon and power counting for quantities we need to establish the relation \refb{lim2}. 

The main aim is to derive the leading term of Taylor expansion of the scalar
\begin{equation}
\left(\tilde{X}^{(12)}_{abc}\nabla_d E^{abcd}\right).
\end{equation} 
For that purpose we need to describe how to count the powers of various quantities. In particular,
leading power of Taylor expansion of some scalar will be called order of that scalar.
Having that in mind and also having in mind relations \refb{CT1}, we can give definitions for the order of basis vectors as
\begin{equation}
\ord(\cchi)=1\,, \quad \ord(\ttt)=0
\end{equation} 
For arbitrary tensor $T$ we first expand it in basis ${e_\mu}^a$: 
\begin{widetext}
\begin{equation}
{T
^{\idxrangeb{a}{1}{m}}}
_{\idxrangeb{b}{1}{n}}
 = \sum_{
\idxrangeb{\mu}{1}{m},
\idxrangeb{\nu}{1}{n}}
T
^{\idxrangeb{\mu}{1}{m},\idxrangeb{\nu}{1}{n}}
{e_{\mu_1}}^{a_1}\ldots {e_{\mu_m}}^{a_m}
e_{\nu_1a_1}\ldots e_{\nu_nb_n}.
\end{equation} 
Then we calculate order for each term in the sum as sum of orders of its factors. 
The order of $T$ is defined as the order of its leading term (i.e.\ of the term with the lowest order):
\begin{equation}\label{defoT}
\ord({T
^{\idxrangeb{a}{1}{m}}}
_{\idxrangeb{b}{1}{n}}
) = \min_{
\idxrangeb{\mu}{1}{m},
\idxrangeb{\nu}{1}{n}}
\left[
\ord(T
^{\idxrangeb{\mu}{1}{m},\idxrangeb{\nu}{1}{n}})
+\sum_{i=0}^{m}\ord({e_{\mu_i}}^{a_i})
+\sum_{i=0}^{n}\ord(e_{\nu_ib_i})
\right].
\end{equation} 
\end{widetext}

Note that, by definition, $e_{\mu a}$ and ${e_\mu}^a$ are of the same order. 
The definition \refb{defoT} implies that for basis vectors ${e_\mu}^a$ and ${e_\nu}^b$:
\begin{equation}\label{oeet}
\ord(e_{\mu a} {e_\nu}^b) = \ord(e_{\mu a}) + \ord({e_\nu}^b)
\end{equation}
and when we contract indices we get [from \refb{CT1} and \refb{CT2}]
\begin{equation}\label{oeec}
\ord(e_{\mu a} {e_\nu}^a) \geq \ord(e_{\mu a}) + \ord({e_\nu}^a)
\end{equation} 
[For example from \refb{CT1} we have $\ord(\chi\cdot\phi)=2$, while $\ord(\chi)=1$ and $\ord(\phi)=0$.] 
For products of tensors we have an analogous situation. For tensors $T_1$ and $T_2$ we have, of course, $\ord(T_1\otimes T_2) = \ord(T_1) + \ord(T_2)$
[i.e.\ when there are no contractions, the leading term is the tensor product of two leading terms], and also in the case of arbitrary contractions we have from \refb{defoT} and \refb{oeec} that
\begin{equation}\label{ottt}
\ord({T_1}^{\ldots}_{\ldots}\ldots{T_n}^{\ldots}_{\ldots}) \geq 
\sum_{i=1}^n\ord({T_i}^{\ldots}_{\ldots}).
\end{equation}
The right hand side of \refb{ottt} gives lower bound for order of arbitrary product of tensors ${T_1}^{\ldots}_{\ldots}\ldots{T_n}^{\ldots}_{\ldots}$ (with possible arbitrary contractions of indices) which is suitable for our purpose of showing \refb{odEXt}.

The fact that $e_{\mu a}$ and ${e_\mu}^a$ are of the same order is consistent with \refb{ottt} and the fact that $g_{ab}$ and $g^{ab}$ are of order $0$.

Important role will have
\begin{eqnarray}
\nonumber
\nabla_a \chi_b &=& \frac{1}{\kappa n^2}(-\chi_a \rho_b + \rho_a \chi_b) +
\textrm{order $\geq 1$ terms}\\\nonumber
\nabla_a \rho_b &=& \frac{1}{\kappa n^2}(\rho_a \rho_b - \chi_a \chi_b) +
\textrm{order $\geq 1$ terms}\\\nonumber
\nabla_a \phi_b &=& \frac{A(z)}{n^2}(-\chi_a \rho_b + \rho_a \chi_b) \\\nonumber
 &+& B(z)(-\phi_a z_b+z_a \phi_b) +
\textrm{order $\geq 1$ terms}\\\nonumber
\nabla_a z_b &=& C(z)(\phi_a \phi_b + z_a z_b) +
\textrm{order $\geq 1$ terms},
\end{eqnarray} 
and we can summarize them as
\begin{eqnarray}\label{DVT}
\nabla_a {e_{\h b}} &\sim & \frac{1}{n^2} {e_{\h a}}{e_{\h b}}
+\textrm{order $\geq 1$ terms}\\\nonumber
\nabla_a {e_{\t b}} &\sim & 
\frac{1}{n^2} {e_{\h a}}{e_{\h b}} + {e_{\t a}}{e_{\t b}}
+\textrm{order $\geq 1$ terms}.
\end{eqnarray} 
Also
\begin{eqnarray}\label{DC}
\nabla_a t &=& -\frac{1}{\kappa^2 n^2}\chi_a + 
\textrm{order $\geq 0$ terms}\\\nonumber
\nabla_a n &=& \frac{1}{\kappa n}\rho_a + \textrm{order $\geq 1$ terms}\\\nonumber
\nabla_a \phi &=& -\frac{\Omega_H}{\kappa^2 n^2}\chi_a + \textrm{order $\geq 0$ terms}\\\nonumber
\nabla_a z &=& \frac{1}{g_{Hzz}(z)}z_a + \textrm{order $\geq 1$ terms}.
\end{eqnarray}

Derivative lowers the order \refb{defoT} at most by one. That can be seen from \refb{DC} and \refb{DVT}
\begin{equation}\label{odT}
\ord(\nabla T) \geq \ord(T) - 1.
\end{equation} 
For a function $f(z)$ it follows from \refb{DC} that 
\begin{equation}\label{DFZ}
\nabla_a f(z) = \frac{\partial f}{\partial z}\frac{1}{g_{Hzz}(z)}z_a +
\textrm{nonleading terms},
\end{equation} 
so that in this case 
\begin{equation}
\ord(\nabla_a f(z)) \geq \ord(f(z)).
\end{equation} 
From \refb{CT1} and \refb{CT2} we see that
${e_\mu}\cdot\chi = O(n^2)$ for $\mu=1,2$ and 
${e_\mu}\cdot\chi = 0$ for $\mu=3,4$, so we can write
\begin{equation}
\ord({e_\mu}\cdot\chi) \geq 2,
\end{equation} 
where, since we are interested in Taylor expansion around $n=0$, we can formally treat $0$ as $O(n^\infty)$, and that is why there is $\geq$ sign. There is analoguos relation for $\rho$
\begin{equation}
\ord({e_\mu}\cdot\rho) \geq 2,
\end{equation} 
so we can write
\begin{equation}\label{r1b}
\ord({e_\mu}\cdot\cchi) \geq 2.
\end{equation} 
We also note that following relations hold:
\begin{equation}\label{r2}
\ord(\nabla_a \ttt_b) \geq 0 ,
\end{equation} 
\begin{equation}\label{r3}
\ord(\nabla_a (\frac{1}{n^2} \chi_{[b}\rho_{c]})) > -1.
\end{equation} 
These relations will enable us later to raise the lower bound calculated by the right hand side of \refb{ottt}.

Since Lagrangian is of the form \refb{lgen}, it can be expressed as a function of scalar invariants $I_n$ 
\begin{equation}
L=\f(I_1,I_2,\ldots),
\end{equation} 
(e.g.\ we can take 
$I_1 = R$, $I_2 = R_{abcd}R^{abcd}$, $I_3 = R_{ab}R^{ab}$, $I_4 = R^2$, \ldots).
Since $L$ does not contain derivatives of Riemann tensors $E^{abcd}$ is given by \refb{e7}
\begin{equation}\label{defE}
E^{abcd} = \frac{\partial L}{\partial R_{abcd}} 
= \frac{\partial \f}{\partial I_n}\frac{\partial I_n}{\partial R_{abcd}} 
\equiv \frac{\partial \f}{\partial I_n}E_{I_n}^{abcd}.
\end{equation} 
Since $\ord(g_{ab}) = 0$ and $\ord(R_{abcd}) = 0$ [by explicit calculation, see \refb{appg} and \refb{appr}], from \refb{ottt} it follows that for $E_{I_n}^{abcd}$ defined in Eq.~\refb{defE} we have
\begin{equation}
\ord(E_{I_n}^{abcd}) \geq 0
\end{equation} 
because $I_n$ consists only of tensors $R_{abcd}$ and $g_{ab}$.
If in addition we require that for each scalar invariant $I_n$
\begin{equation}\label{lagreg}
\frac{\partial \f}{\partial I_n} = \textrm{finite on the horizon},
\end{equation} 
then \refb{lagreg} implies that
\begin{equation}\label{oE}
\ord(E^{abcd}) \geq 0
\end{equation}

Now we write $E^{abcd}$ using components $E^{\mu\nu\rho\sigma}$ in basis ${e_\mu}^a$ 
\begin{equation}
E^{abcd} = \sum_{\mu\nu\rho\sigma}E^{\mu\nu\rho\sigma}
{e_\mu}^a
{e_\nu}^b
{e_\rho}^c
{e_\sigma}^d.
\end{equation} 
In the same way we also expand derivative $\nabla^eE^{abcd}$
\begin{equation}
\nabla^eE^{abcd} = \sum_{\mu\nu\rho\sigma\lambda}K^{\mu\nu\rho\sigma\lambda}
{e_\mu}^a
{e_\nu}^b
{e_\rho}^c
{e_\sigma}^d
{e_\lambda}^e,
\end{equation} 
and contraction
\begin{equation}\label{bdE}
\nabla_dE^{abcd} = \sum_{\mu\nu\rho}C^{\mu\nu\rho}
{e_\mu}^a
{e_\nu}^b
{e_\rho}^c.
\end{equation} 
Note that components $E^{\mu\nu\rho\sigma}$, 
$K^{\mu\nu\rho\sigma\lambda}$ and 
$C^{\mu\nu\rho}$ have symmetries which follow from symmetries of Riemann tensor, and also, in this basis, these components are functions of $n$ and $z$ only.

From \refb{oE} and \refb{odT} we have for each $\mu$, $\nu$, $\rho$, $\sigma$, $\lambda$
\begin{equation}
\ord(E^{\mu\nu\rho\sigma}
{e_\mu}^a
{e_\nu}^b
{e_\rho}^c
{e_\sigma}^d)\geq 0,
\end{equation} 
\begin{equation}
\ord(K^{\mu\nu\rho\sigma\lambda}
{e_\mu}^a
{e_\nu}^b
{e_\rho}^c
{e_\sigma}^d
{e_\lambda}^e)\geq -1,
\end{equation} 
\begin{equation}
\ord(C^{\mu\nu\rho}
{e_\mu}^a
{e_\nu}^b
{e_\rho}^c) \geq -1.
\end{equation} 
That implies:
\begin{eqnarray}\label{tE}
\ord(E^{\h\h\h\h}) &\geq & -4\\\nonumber
\ord(E^{\h\h\h\t}) &\geq & -3\\\nonumber
\ord(E^{\h\h\t\t}) &\geq & -2\\\nonumber
\textrm{etc.}&&
\end{eqnarray} 
\begin{eqnarray}\label{tK}
\ord(K^{\h\h\h\h\h}) &\geq & -6\\\nonumber
\ord(K^{\h\h\h\h\t}) &\geq & -5\\\nonumber
\ord(K^{\h\h\h\t\h}) &\geq & -5\\\nonumber
\ord(K^{\h\h\h\t\t}) &\geq & -4\\\nonumber
\textrm{etc.}&&
\end{eqnarray} 
\begin{eqnarray}\label{tC}
\ord(C^{\h\h\h}) \geq -4\\\nonumber
\ord(C^{\h\h\t}) \geq -3\\\nonumber
\ord(C^{\h\t\h}) \geq -3\\\nonumber
\ord(C^{\h\t\t}) \geq -2\\\nonumber
\ord(C^{\t\t\h}) \geq -2\\\nonumber
\ord(C^{\t\t\t}) \geq -1\\\nonumber
\end{eqnarray} 
The coefficients $K^{\mu\nu\rho\sigma\lambda}$ and $E^{\mu\nu\rho\sigma}$ are related with
\begin{eqnarray}\label{rKE}
&&\sum_{\mu\nu\rho\sigma\lambda}K^{\mu\nu\rho\sigma\lambda}
{e_\mu}^a
{e_\nu}^b
{e_\rho}^c
{e_\sigma}^d
{e_\lambda}^e = \\\nonumber
&&\qquad\nabla^e \sum_{\mu\nu\rho\sigma}\left(E^{\mu\nu\rho\sigma}
{e_\mu}^a
{e_\nu}^b
{e_\rho}^c
{e_\sigma}^d\right).
\end{eqnarray} 
The coefficients $C^{\mu\nu\rho}$ and $K^{\mu\nu\rho\sigma\lambda}$ are related with
\begin{equation}\label{rCK}
C^{\mu\nu\rho} = \sum_{\sigma\lambda}K^{\mu\nu\rho\sigma\lambda}{e_\sigma}\cdot{e_\lambda}.
\end{equation}

Now we prove that \refb{lim2} holds
for Lagrangians of type \refb{lgen}. That will be the case if 
\begin{equation}\label{odEXt}
\ord(\nabla_dE^{abcd}\tilde{X}^{(12)}_{abc}) > 0.
\end{equation} 

Explicit calculation \refb{appx} of $\tilde{X}^{(12)}_{abc}$  (whose form does not depend on Lagrangian) tells us that the leading terms in it are of order $1$:
\begin{equation}\label{oXt}
\ord(\tilde{X}^{(12)}_{abc}) = 1,
\end{equation}
and these are
\begin{eqnarray}\label{lXt}
&&\chi_a\rho_b\chi_cO(\frac{1}{n^2})+
\chi_a\rho_b\rho_cO(\frac{1}{n^2})+\\\nonumber
&&\quad+\rho_a\chi_b\chi_cO(\frac{1}{n^2})+
\rho_a\chi_b\rho_cO(\frac{1}{n^2}),
\end{eqnarray}
and also that 
\begin{equation}\label{ooXt}
\ord(\textrm{other terms in $\tilde{X}^{(12)}_{abc}$}) \geq 2.
\end{equation}  
On the other hand we see from \refb{oE} and \refb{odT} that
\begin{equation}\label{odE}
\ord(\nabla_dE^{abcd}) \geq -1.
\end{equation}
If we show that leading terms (of order $-1$) in $\nabla_dE^{abcd}$ cancel when contracted with leading terms \refb{lXt} (of  order $1$) in $\tilde{X}^{(12)}_{abc}$, then \refb{odEXt} will follow.

We look at terms \refb{lXt} contracted with \refb{bdE} and using \refb{ottt} we count the order to be at least $0$:
\begin{eqnarray}
&& \ord\left[\sum_{\mu\nu\lambda}C^{\mu\nu\lambda} \left(
{e_\mu} \!\! \cdot \!\! \chi \;
{e_\nu} \!\! \cdot \!\! \rho \;
{e_\lambda} \!\! \cdot \!\! \chi \; O(\frac{1}{n^2})
\right.
\right.
\\\nonumber 
&&\hspace{2.5cm}+
\left.
\left.
{e_\mu} \!\! \cdot \!\! \chi \;
{e_\nu} \!\! \cdot \!\! \rho \;
{e_\lambda} \!\! \cdot \!\! \rho \; O(\frac{1}{n^2})
\right)\right] \geq 0,
\end{eqnarray} 
where we used $C^{\mu\nu\rho} = C^{[\mu\nu]\rho}$.
To prove \refb{odEXt} we need to prove
\begin{eqnarray}\label{e66}
&&\ord\left[C^{\mu\nu\lambda} \left(
{e_\mu} \!\! \cdot \!\! \chi \;
{e_\nu} \!\! \cdot \!\! \rho \;
{e_\lambda} \!\! \cdot \!\! \chi \; O(\frac{1}{n^2})
\right.
\right.
\\\nonumber 
&&\hspace{2cm}+
\left.
\left.
{e_\mu} \!\! \cdot \!\! \chi \;
{e_\nu} \!\! \cdot \!\! \rho \;
{e_\lambda} \!\! \cdot \!\! \rho \; O(\frac{1}{n^2})
\right)\right] > 0,
\end{eqnarray} 
for each $\mu$, $\nu$ and $\lambda$.
Using \refb{r1b} and \refb{tC} we see that \refb{e66} will follow if 
\begin{equation}\label{e39}
\ord(C^{\h\h\h}) > -4.
\end{equation} 
On the other hand from \refb{ottt} and \refb{rCK} we see that
\begin{equation}
\ord(C^{\h\h\h}) \geq \ord\left(
\sum_{\lambda\sigma}K^{\h\h\h\lambda\sigma}{e_\lambda}\cdot{e_\sigma}
\right).
\end{equation} 
Writing in terms of the lower bound of right hand side we get
\begin{equation}
\ord(C^{\h\h\h}) \geq 
\min_{\lambda\sigma}\ord\left(K^{\h\h\h\lambda\sigma}{e_\lambda}\cdot{e_\sigma}
\right).
\end{equation} 
Expanding $\lambda$ and $\sigma$ we get
\begin{eqnarray}
\ord(C^{\h\h\h})  \geq 
\min\left\lbrace      \vphantom{\left.\ord\left(K^{\h\h\h\t\t}{e_{\t}}\cdot{e_{\t}}\right)\right\rbrace}
\right.
&&\ord{}\left(K^{\h\h\h\h\h}{e_{\h}}\cdot {e_{\h}}
\vphantom{K^{\h\h\h\t\t}{e_{\t}}\cdot{e_{\t}}}\right),\\\nonumber
&&\ord{}\left(K^{\h\h\h\h\t}{e_{\h}}\cdot {e_{\t}}\right),\\\nonumber
&&\ord{}\left(K^{\h\h\h\t\h}{e_{\t}}\cdot {e_{\h}}\right),\\\nonumber
&&\left.\ord{}\left(K^{\h\h\h\t\t}{e_{\t}}\cdot{e_{\t}}\right)\right\rbrace
\end{eqnarray} 
Explicitly
\begin{eqnarray}
&&\ord(C^{\h\h\h}) \\\nonumber
&& \qquad \geq  \min\left\lbrace     (-6)+2, (-5)+2, (-5)+2, (-4)+0 \right\rbrace \\\nonumber
&& \qquad \geq  \min\left\lbrace     -4, -3, -3, -4 \right\rbrace
\end{eqnarray} 
So if we prove that 
\begin{equation}\label{K5}
\ord(K^{\h\h\h\h\h}) > -6
\end{equation} 
and
\begin{equation}\label{K3}
\ord(K^{\h\h\h\t\t}) > -4
\end{equation} 
then \refb{e39} will hold.

From \refb{DVT} and \refb{rKE} we see that $K^{\h\h\h\h\h}$ can only get contribution from $4$
terms in the sum on the right hand side of \refb{rKE} which are of the form
\begin{equation}\label{ee1}
E^{\h\h\h\h}{e_{\h}}^a{e_{\h}}^b{e_{\h}}^c{e_{\h}}^d
\end{equation} 
and $4\cdot 8=32$ terms of the form
\begin{eqnarray}\label{ee2}
E^{\t\h\h\h}{e_{\t}}^a{e_{\h}}^b{e_{\h}}^c{e_{\h}}^d\\\nonumber
E^{\h\t\h\h}{e_{\h}}^a{e_{\t}}^b{e_{\h}}^c{e_{\h}}^d\\\nonumber
E^{\h\h\t\h}{e_{\h}}^a{e_{\h}}^b{e_{\t}}^c{e_{\h}}^d\\\nonumber
E^{\h\h\h\t}{e_{\h}}^a{e_{\h}}^b{e_{\h}}^c{e_{\t}}^d\\\nonumber
\end{eqnarray} 
when derivative acts as
\begin{equation}\label{eee1}
(\nabla_eE^{\h\h\h\h}){e_{\h}}^a{e_{\h}}^b{e_{\h}}^c{e_{\h}}^d
\end{equation} 
or as
\begin{eqnarray}\label{eee2}
E^{\h\h\h\h}(\nabla_e{e_{\h}}^a){e_{\h}}^b{e_{\h}}^c{e_{\h}}^d\\\nonumber
E^{\h\h\h\h}{e_{\h}}^a(\nabla_e{e_{\h}}^b){e_{\h}}^c{e_{\h}}^d\\\nonumber
E^{\h\h\h\h}{e_{\h}}^a{e_{\h}}^b(\nabla_e{e_{\h}}^c){e_{\h}}^d\\\nonumber
E^{\h\h\h\h}{e_{\h}}^a{e_{\h}}^b{e_{\h}}^c(\nabla_e{e_{\h}}^d)\\\nonumber
\end{eqnarray} 
on \refb{ee1}, and as
\begin{eqnarray}\label{eee3}
E^{\t\h\h\h}(\nabla_e{e_{\t}}^a){e_{\h}}^b{e_{\h}}^c{e_{\h}}^d\\\nonumber 
E^{\h\t\h\h}{e_{\h}}^a(\nabla_e{e_{\t}}^b){e_{\h}}^c{e_{\h}}^d\\\nonumber
E^{\h\h\t\h}{e_{\h}}^a{e_{\h}}^b(\nabla_e{e_{\t}}^c){e_{\h}}^d\\\nonumber
E^{\h\h\h\t}{e_{\h}}^a{e_{\h}}^b{e_{\h}}^c(\nabla_e{e_{\t}}^d)\\\nonumber
\end{eqnarray} 
on \refb{ee2}.
And also we see that $K^{\h\h\h\t\t}$ can only get contribution from $8$
terms of the form
\begin{equation}
E^{\h\h\h\t}{e_{\h}}^a{e_{\h}}^b{e_{\h}}^c{e_{\t}}^d
\end{equation} 
when derivative acts as
\begin{equation}
E^{\h\h\h\t}{e_{\h}}^a{e_{\h}}^b{e_{\h}}^c(\nabla_e{e_{\t}}^d)
\end{equation} 
which is included in \refb{eee3}.

The sum of \refb{eee1} and \refb{eee2} is
\begin{eqnarray}\label{eee12}
&& \nabla_e(E^{\h\h\h\h}{e_{\h}}^a{e_{\h}}^b{e_{\h}}^c{e_{\h}}^d) \\\nonumber
&&\qquad = 4\nabla_e(E^{1212}(n,z){e_{[1}}^a{e_{2]}}^b{e_{[1}}^c{e_{2]}}^d)  \\\nonumber
&&\qquad= 4\nabla_e(\frac{f(z)}{n^4}{e_{[1}}^a{e_{2]}}^b{e_{[1}}^c{e_{2]}}^d + 
\textrm{order $> 0$ terms}) \\\nonumber
\end{eqnarray} 
where 
$f(z) \equiv \lim_{n\to 0}[n^4 E^{1212}(n,z)]$, and has order $\geq 0$ because of \refb{tE}. The leading term in \refb{eee12} is
\begin{eqnarray}\label{eee122}
&& \nabla_e(\frac{f(z)}{n^4}{e_{[1}}^a{e_{2]}}^b{e_{[1}}^c{e_{2]}}^d) \\\nonumber
&&\qquad = (\nabla_ef(z))\frac{1}{n^4}{e_{[1}}^a{e_{2]}}^b{e_{[1}}^c{e_{2]}}^d \\\nonumber
&&\qquad \hphantom{=}+ f(z)(\nabla_e\frac{1}{n^2}{e_{[1}}^a{e_{2]}}^b)(\frac{1}{n^2}{e_{[1}}^c{e_{2]}}^d) \\\nonumber
&&\qquad \hphantom{=}+ f(z)(\frac{1}{n^2}{e_{[1}}^a{e_{2]}}^b)(\nabla_e\frac{1}{n^2}{e_{[1}}^c{e_{2]}}^d) \\\nonumber
\end{eqnarray} 
Using \refb{DFZ}, \refb{r3} and \refb{ottt} and inserting \refb{eee122} into \refb{eee12} we get 
\begin{equation}
\ord(\nabla_e(E^{\h\h\h\h}{e_{\h}}^a{e_{\h}}^b{e_{\h}}^c{e_{\h}}^d)) > -1
\end{equation}

Also from \refb{r2} and \refb{ottt} we get bound for order of Eqs.~\refb{eee3} 
\begin{equation}
\ord(E^{\t\h\h\h}(\nabla_e{e_{\t}}^a){e_{\h}}^b{e_{\h}}^c{e_{\h}}^d) > -1.
\end{equation}
Using \refb{rKE} we see that \refb{K5} and \refb{K3} follow, which completes the proof of \refb{lim2}.
Now we are able to use properties \refb{lim1} and \refb{lim2} to calculate the central charge
\begin{equation}
\frac{c}{12} = \frac{\hat{A}}{8\pi},
\end{equation}
and entropy formula \refb{e60}. The last derivation is analogous to one in Refs.~\cite{Carlip:1999cy,Cvitan:2003vq}.

\section{Conclusion}
Derivation of black hole entropy \cite{Carlip:1999cy,Cvitan:2002rh,Cvitan:2003vq} which used ideas of conformal symmetry and Virasoro algebras have been based on additional plausible assumptions. It is important to find examples of theories where this assumptions are fulfilled. It is also important to understand if they depend on properties of interactions or instead on the properties of horizons. In this paper we show that latter is the case. In fact using the properties of horizons of stationary black hole which follow from regularity of curvature invariants, one can derive the mentioned result. This was done for Einstein gravity and quadratic Lagrangians in the previous reference \cite{Cvitan:2004cj}
by explicit calculation. This is not possible for generic case and thus we have used here a new method based on power counting. In such a way we have been able to generalise rhe result to Lagrangians with arbitrary dependence on Riemann tensor.

In particular, inverse powers of invariants are allowed terms. They are restricted with condition \refb{lagreg}. As mentioned in the introduction such Lagrangians are also of interest due to the present effort to investigate if they could accomodate acceleration of the universe.
Of course an investigation valid for Lagrangians involving derivatives of Riemann tensor is still missing. However, in the Appendix \ref{appb} we present a special case of such 
Lagrangian where these results are again true. As a byproduct of this investigation is the conclusion that requiring the regularity of invariants involving derivatives of Riemann tensor gives even more restrictions on metric functions.

\begin{acknowledgments}
We would like to acknowledge the financial support under the contract
No.\ 0119261 of 
Ministry of Science, Education and Sports
%Ministry of Science and Technology 
of Republic of Croatia.
\end{acknowledgments}

\appendix

\section{}
In this text we have used relation \refb{oE} stating that
\begin{equation}\label{aoE}
\ord(E^{abcd}) \geq 0
\end{equation} 
That relation was in turn consequence of properties
\begin{equation}\label{aogR}
\ord(g^{ab}) = 0\;,\quad
\ord(R_{abcd}) = 0
\end{equation} 
which are consequence of Taylor expansions \refb{mt} for metric functions. We have also used relation
\refb{oXt}
\begin{equation}\label{aoXt}
\ord(\tilde{X}^{(12)}_{abc}) = 1,
\end{equation}
based also on \refb{mt}.
In this Appendix we  give decomposition of tensors $g_{ab}$, $\tilde{X}^{(12)}_{abc}$  and $R_{abcd}$ in the basis $\left\lbrace \chi^a, \rho^a, \phi^a, z^a\right\rbrace$ introduced in text. These decompositions are result of lengthy but straightforward calculations which can be done e.g.\ with the help of Mathematica. We give these expressions:
\begin{eqnarray}\label{appg}
g_{ab} &=& \chi_a\chi_b\left( -\frac{1}{\kappa^2 n^2} + O(n^{-1})\right)\\\nonumber
&+&\rho_a\rho_b\left( \frac{1}{\kappa^2 n^2} + O(n^{-1})\right)\\\nonumber
&+&\phi_a\phi_b\left( \frac{1}{g_{H\phi\phi}} + O(n)\right)\\\nonumber
&+&z_az_b\left( \frac{1}{g_{Hzz}} + O(n)\right)\\\nonumber
&+& \textrm{terms of order $\geq 1$}
\end{eqnarray}
\begin{eqnarray}\label{appx}
\tilde{X}_{abc}^{(12)}&=&{{\chi }_a}{{\rho }_b}{{\chi }_c}\left(
-\frac{
{{{\dot{T}}}_1}{}{{{\ddot{T}}}_2} -
{{{\dot{T}}}_2}{}{{{\ddot{T}}}_1}  
}
{2{}{\kappa }^4{}n^2} + {O}(n^{-1})
\right)\\\nonumber
&+&{{\chi }_a}{{\rho }_b}{{\rho }_c}\left(
\frac{
{{{\ddot{T}}}_1}{}{T_2}-
{{{\ddot{T}}}_2}{}{T_1}  
}
{2{}{\kappa }^3{}n^2} + {O}(n^{-1})\right)\\\nonumber
&+&{{\rho }_a}{{\chi }_b}{{\chi }_c}\left(
\frac{
{{{\dot{T}}}_1}{}{{{\ddot{T}}}_2} -
{{{\dot{T}}}_2}{}{{{\ddot{T}}}_1}  
}
{2{}{\kappa }^4{}n^2} + {O}(n^{-1})\right)\\\nonumber
&+&{{\rho }_a}{{\chi }_b}{{\rho }_c}\left(-\frac{
{{{\ddot{T}}}_1}{}{T_2}-
{{{\ddot{T}}}_2}{}{T_1} 
}
{2{}{\kappa }^3{}n^2} + {O}(n^{-1})\right)\\\nonumber
&+& \textrm{terms of order $\geq 2$}
\end{eqnarray} 
\begin{widetext}
\begin{eqnarray}\label{appr}
R_{abcd}
&=&{{\chi }_a}{{\rho }_b}{{\chi }_c}{{\rho }_d}\left(
-\frac
{R_{\h} }
{2{}{\kappa }^4{}n^4} 
+ {{O}(n^{-3})}
\right)
 + {{\chi }_a}{{\rho }_b}{{\rho }_c}{{\phi }_d}\left(\frac{3{}\omega_{3}}{2{}{\kappa }^4{}n^3} + {{O}(n^{-2})}\right)\\\nonumber
&+&{{\chi }_a}{{\rho }_b}{{\phi }_c}{z_d}\left(
-\frac{g_{H\phi\phi}{}\omega_{2}' + \omega_{2}{}g_{H\phi\phi}'}
{2{}g_{H\phi\phi}{}g_{Hzz}{}{\kappa }^3{}n^2} + {O}(n^{-1})\right)
 + {{\chi }_a}{{\phi }_b}{{\chi }_c}{{\phi }_d}\left(\frac{{\omega_{2}}^2 + \frac{2{}g_{2\phi\phi}{}{\kappa }^2}{{g_{H\phi\phi}}^2}}{4{}{\kappa }^4{}n^2} + {O}(n^{-1})\right)\\\nonumber
&+&{{\chi }_a}{{\phi }_b}{{\rho }_c}{z_d}\left(
-\frac{g_{H\phi\phi}{}\omega_{2}' + \omega_{2}{}g_{H\phi\phi}'}
{4{}g_{H\phi\phi}{}g_{Hzz}{}{\kappa }^3{}n^2} 
+ {O}(n^{-1})\right)
 + {{\chi }_a}{z_b}{{\chi }_c}{z_d}\left(\frac{g_{2zz}}{2{}{g_{Hzz}}^2{}{\kappa }^2{}n^2} + {O}(n^{-1})\right)\\\nonumber
&+&{{\chi }_a}{z_b}{{\rho }_c}{{\phi }_d}\left(
\frac{g_{H\phi\phi}{}\omega_{2}' + \omega_{2}{}g_{H\phi\phi}'}
{4{}g_{H\phi\phi}{}g_{Hzz}{}{\kappa }^3{}n^2} + {O}(n^{-1})\right)
 + {{\rho }_a}{{\phi }_b}{{\rho }_c}{{\phi }_d}\left(
-\frac{ {\omega_{2}}^2 + \frac{2{}g_{2\phi\phi}{}{\kappa }^2}{{g_{H\phi\phi}}^2} }
{4{}{\kappa }^4{}n^2} + {O}(n^{-1})\right)\\\nonumber
&+&{{\rho }_a}{z_b}{{\rho }_c}{z_d}\left(\frac{-g_{2zz}}{2{}{g_{Hzz}}^2{}{\kappa }^2{}n^2} + {O}(n^{-1})\right)
 + {{\phi }_a}{z_b}{{\phi }_c}{z_d}\left(
\frac{R_{\t}}
{2{}{g_{H\phi\phi}}{}{g_{Hzz}}} + {O}(n)\right)\\\nonumber
&+& \textrm{terms of order $\geq 1$}\\\nonumber
&+& \textrm{terms related by symmetries of $R_{abcd}$},
\end{eqnarray}
\end{widetext}
where $\omega_3$ is defined as $\omega(n,z) = \Omega_H + {1\over 2} \omega_2(z) n^2 +  \omega_3(z) n^3 + O(n^4)$, and 
\begin{equation}
R_{\h} = 
\frac
{3{}{\omega_{2}}^2{}g_{H\phi\phi} - 4{}\kappa_2{}\kappa }
{2{\kappa }^2} ,
\end{equation} 
\begin{equation}
R_{\t} = \frac
{g_{Hzz}{}{g_{H\phi\phi}'}^2 + g_{H\phi\phi}{}g_{H\phi\phi}'{}g_{Hzz}' - 2{}g_{H\phi\phi}{}g_{Hzz}{}g_{H\phi\phi}''}
{2{}{g_{H\phi\phi}}^2{}{g_{Hzz}}^2}.
\end{equation} 
From these expressions the properties \refb{aoE} and \refb{aoXt} can be read.
 
\section{}\label{appb}
An analysis which would include generic Lagrangians involving derivatives of Riemann tensor is of course much more complex. Here we consider a special case where we add to Lagrangians \refb{lgen} the term
\begin{equation}\label{eqaddition}
(\nabla_a R)^2.
\end{equation} 
Now, Dirac brackets 
\begin{equation}
\left\lbrace J[\xi_1],J[\xi_2]\right\rbrace ^* = 
\int_{\mathcal{H}} \left(
\xi_2\cdot\theta_1
   - \xi_1\cdot\theta_2
   - \xi_2\cdot(\xi_1\cdot{\mathbf{L}}) \right)
\end{equation} 
change by the term
\begin{eqnarray}\nonumber
\int_{\mathcal{H}} &\hat{\bm{\epsilon}}&
\left\lbrace 2\left(
  X^{(12)}_{abcd}E^{abcd}
  -\tilde{X}^{(12)}_{abc}F^{abc}
\right)\right\rbrace
\end{eqnarray}
where 
\begin{equation}
E^{abcd} = \frac{1}{2}\left(g^{ad}g^{bc} - g^{ac}g^{bd}\right)\nabla^2R
\end{equation} 
and
\begin{eqnarray}\nonumber
F^{abc} &=& 
\nabla^b\nabla^c\nabla^aR 
- 2g^{bc}\nabla^a\nabla^2R
+ g^{ac}\nabla^b\nabla^2R\\
&-&g^{bc}R^{ae}\nabla_eR \\\nonumber
&-&R^{bc}\nabla^aR
+2R^{ac}\nabla^bR
-2R^{ab}\nabla^cR
\end{eqnarray} 

A long but straightforward calculation shows that for special case \refb{eqaddition} usual results can be obtained provided we restrict the class of metric functions. The restrictions are
\begin{equation}\label{restr1}
\omega_3 = 0
\end{equation} 
and
\begin{equation}\label{restr2}
\frac{3 g_{3zz}}{g_{Hzz}} + 
\frac{8 \kappa_{3}}{\kappa} + 
\frac{3 g_{3\phi\phi}}{g_{H\phi\phi}} = 0
\end{equation} 
where $g_{3zz}(z)$, $g_{3\phi\phi}(z)$ and $\omega_3(z)$ are coefficients of $n^3$ , and $\kappa_{3}(z)$ of $n^4$ in Taylor expansions \refb{mt}.

These restrictions can be understood also by terms of regularity of scalar curvature invariants on horizon. Namely, if we require regularity of
\begin{equation}
(\nabla_a R_{bc})^2\qquad\textrm{and}\qquad\nabla^2R
\end{equation} 
we obtain relations \refb{restr1} and \refb{restr2}.

From \refb{restr1} it follows that $\ord(\nabla_eR_{abcd})=0$ and so all polynomial invariants involving Riemann tensor and its first derivatives will be regular on the horizon.
This is in fact generalisation of results from \cite{Medved:2004tp} that regularity of invariants of Riemann tensor has implications on metric functions near horizon. Here, we see that regularity of invariants involving derivatives of Riemann tensor has even stronger consequences on metric functions.

\end{document}